\documentclass[nofootinbib,floats,aps,superscriptaddress,preprintnumbers]{revtex4}

\usepackage{graphicx}
\usepackage[hypertex]{hyperref}

\newcommand{\beq}{\begin{eqnarray}}
\newcommand{\eeq}{\end{eqnarray}}
\newcommand{\nn}{\nonumber}

\def\OMIT#1{}%
\def\d{\mathrm{d}}
\def\lqcd{\Lambda_{\rm QCD}}

\def\ov{\overline}

\newcommand{\tev}{\,{\rm TeV}}
\newcommand{\gev}{\,{\rm GeV}}

\newcommand{\ollu}{O_{LL}^u}
\newcommand{\ollh}{O_{LL}^h}
\newcommand{\olrw}{O_{LR}^w}
\newcommand{\orlw}{O_{RL}^w}
\newcommand{\olrb}{O_{LR}^b}
\newcommand{\orlb}{O_{RL}^b}
\newcommand{\orru}{O_{RR}^u}

\newcommand{\cllu}{C_{LL}^u}
\newcommand{\cllh}{C_{LL}^h}
\newcommand{\clrw}{C_{LR}^w}
\newcommand{\crlw}{C_{RL}^w}
\newcommand{\clrb}{C_{LR}^b}
\newcommand{\crlb}{C_{RL}^b}
\newcommand{\crru}{C_{RR}^u}

\newcommand{\ollup}{O_{LL}^{\prime u}}
\newcommand{\ollhp}{O_{LL}^{\prime h}}
\newcommand{\olrwp}{O_{LR}^{\prime w}}
\newcommand{\orlwp}{O_{RL}^{\prime w}}
\newcommand{\olrbp}{O_{LR}^{\prime b}}

\newcommand{\cllup}{C_{LL}^{\prime u}}
\newcommand{\cllhp}{C_{LL}^{\prime h}}
\newcommand{\clrwp}{C_{LR}^{\prime w}}
\newcommand{\crlwp}{C_{RL}^{\prime w}}
\newcommand{\clrbp}{C_{LR}^{\prime b}}
\newcommand{\crlbp}{C_{RL}^{\prime b}}
\newcommand{\crrup}{C_{RR}^{\prime u}}

\newcommand{\lnorm}{ { 1\, \mathrm{TeV} } }

\newcommand{\qor}{ \quad \mathrm{or} \quad}

\begin{document}

\pagestyle{plain}  

\preprint{\vbox{\hbox{UCB--PTH--07/06}\hbox{YITP--SB--07--11}}}

\title{\boldmath Deciphering top flavor violation at the LHC with $B$ factories}

\author{Patrick J.\ Fox}
\affiliation{Ernest Orlando Lawrence Berkeley National Laboratory,
University of California, Berkeley, CA 94720}

\author{Zoltan Ligeti}
\affiliation{Ernest Orlando Lawrence Berkeley National Laboratory,
University of California, Berkeley, CA 94720}

\author{Michele Papucci}
\affiliation{Ernest Orlando Lawrence Berkeley National Laboratory,
University of California, Berkeley, CA 94720}
\affiliation{Department of Physics, University of California,
Berkeley, CA 94720}

\author{Gilad Perez}
\affiliation{C.N. Yang Institute for Theoretical Physics, 
State University of New York, Stony Brook, NY 11794-3840}

\author{Matthew D. Schwartz}
\affiliation{Department of Physics and Astronomy, Johns Hopkins University,
3400 North Charles St., Baltimore, MD 21218}


\begin{abstract}

The LHC will have unprecedented sensitivity to flavor-changing neutral current
(FCNC) top quark decays, whose observation would be a clear sign of physics
beyond the standard model.  Although many details of top flavor violation are
model dependent, the standard model gauge symmetries relate top FCNCs to other
processes, which are strongly constrained by existing data.  We study these
constraints in a model independent way, using a low energy effective theory from
which the new physics is integrated out.  We consider the most important
operators which contribute to top FCNCs and analyze the current constraints on
them.  We find that the data rule out top FCNCs at a level observable at the LHC
due to most of the operators comprising left-handed first or second generation
quark fields, while there remains a substantial window for top decays mediated
by operators with right-handed charm or up quarks.  If FCNC top decays are
observed at the LHC, such an analysis may help decipher the underlying physics.

\end{abstract}
\maketitle

\section{Introduction}

The Large Hadron Collider (LHC) will have unprecedented sensitivity to flavor
changing neutral currents (FCNCs) involving the top quark, such as $t\to c Z$.
With a $t \ov t$ pair production cross section of about $800$\,pb and after
100\,fb$^{-1}$ of integrated luminosity, the LHC will explore branching ratios
down to the $10^{-5}$ level~\cite{Carvalho:2007yi,cmstdr}.  Flavor changing neutral
currents are highly suppressed in the standard model (SM), but are expected to
be enhanced in many models of new physics (NP). Because top FCNCs are clean
signals, they are a good place to explore new physics. There are important
constraints from $B$ physics on what top decays are allowed, and understanding
these constraints may help decipher such an FCNC signal. In this paper, we
calculate the dominant constraints on top FCNCs from low energy physics and
relate them to the expected LHC reach using a model-independent effective field
theory description.

Flavor physics involving only the first two generations is already highly
constrained, but the third generation could still be significantly affected. Of
course, the new flavor physics could be so suppressed that it will not be
observable at all at the LHC. However, since the stabilization of the Higgs mass
is expected to involve new physics to cancel the top loop, it is natural to
expect some new flavor structure which may show up in the top quark couplings to
other standard model fields.  Thus, one may expect flavor physics to be related
to the  electroweak scale, and then flavor changing effects involving the top
quark are a natural consequence.  

Although there are many models which produce top FCNCs, the low energy
constraints are independent of the details of these models. The new physics can
be integrated out, leaving a handful of operators relevant at the weak scale
involving only standard model fields. These operators mediate both FCNC top
decays and flavor-changing transitions involving lighter quarks. Thus, the two
can be related without reference to a particular model of new physics, provided
there is no additional NP contributing to the $B$ sector. The low energy
constraints can be applied to any model in which top FCNCs are generated and the
constraints on the operators may give information on the scale at which the
physics that generates them should appear.

Analyses of FCNC top decays have been carried out both in the context of
specific models~\cite{modeldep} and using model independent
approaches~\cite{modelindep}.  However, in most cases the effective Lagrangian
analyzed involved the SM fields after electroweak symmetry breaking.  As we
shall see, the scale $\Lambda$ at which the  operators responsible for top FCNC
are generated has to be above the scale $v$ of electroweak symmetry breaking.
Thus, integrating out the new physics should be done before electroweak symmetry
breaking, leading to an operator product expansion in $v/\Lambda$. The
requirement of $SU(2)_L$ invariance provides additional structure on the
effective operators~\cite{Buchmuller:1985jz}, which helps constrain the
expectations for top FCNCs. For example, an operator involving the left-handed
$(t,b)$ doublet, the $SU(2)$ gauge field, and the right handed charm quark, can
lead to $b\to s \gamma$ at one loop, but also directly to a $b\to c$ transition.
If we ignored $SU(2)_L$ invariance, we would only have the $b\to s \gamma$
constraint, and the resulting bound would be different. An important feature of
our analysis is that, after electroweak symmetry breaking, the resulting
operators can modify even SM parameters which contribute at tree level to $B$
physics observables, such as $|V_{cb}|$.

The organization of this paper is as follows. In Sec.~II we introduce the
effective Lagrangian relevant for top FCNCs. We also explain why some operators
can be neglected and introduce conventions used throughout the paper. In
Sec.~III we calculate how these operators affect top quark decays and integrate
out the $W$ and $Z$ bosons and the top quark to match onto the relevant
effective theory at the weak scale.  In Sec.~IV we relate the experimental
constraints to the Wilson coefficients calculated in Section~III, focusing
mostly on observables related to $B$ physics. This leads directly to predictions
for the top branching ratio. Sec.~V contains a summary of the results and our
conclusions. We include an Appendix with details of the calculations.

\section{Effective Lagrangian for top FCNC}
\label{Effective}

We consider an effective Lagrangian
\begin{equation} \label{opnorm}
{\mathcal L}_{\mathrm{eff}} = \frac{1}{\Lambda^2} 
  \sum\, ( C_i\, O_i + C'_i\, O'_i) \,.
\end{equation}
where the $O_i$ operators involve third and second generation quarks and the
$O'_i$ involve the third and first generations. Since we are interested in top
quark decays, we define $O_i$ and $O'_i$ in the mass basis for the up-type
quarks.

A complete set of dimension-six operators which give a $tcZ$ or $tc\gamma$ 
vertex are
\begin{eqnarray}\label{operators}
\ollu &=& i\left[ {\overline Q}_3 {\tilde H} \right]
  \left[ \big( D\!\!\!\!\slash {\tilde H} \big)^\dagger Q_2 \right]
  - i\left[ {\overline Q}_3 \big( D\!\!\!\!\slash {\tilde H} \big) \right]
  \left[ {\tilde H}^\dagger Q_2 \right] + {\mathrm {h.c.}} \,, \nn\\
\ollh &=& i\,\Big[{\overline Q}_3 \gamma^\mu Q_2 \Big] \Big[H^{\dagger}
  \stackrel{\raisebox{-3pt}{$\leftrightarrow$}}{\raisebox{-.5pt}{$D$}}_{\mu}\! H \Big] + {\mathrm {h.c.}} \,, \nn\\
\orlw &=& g_2 \left[ {\overline Q}_2 \sigma^{\mu\nu} \sigma^a \tilde H \right] 
  t_R W_{\mu\nu}^a + {\mathrm {h.c.}} \,, \nn\\
\orlb &=& g_1 \left[{\overline Q}_2 \sigma^{\mu\nu} \tilde H\right] 
  t_R B_{\mu\nu} + {\mathrm {h.c.}} \,, \nn\\
\olrw &=& g_2 \left[{\overline Q}_3 \sigma^{\mu\nu} \sigma^a \tilde H\right] 
  c_R W_{\mu\nu}^a + {\mathrm {h.c.}} \,, \nn\\
\olrb &=& g_1 \left[\overline Q_3 \sigma^{\mu\nu} \tilde H \right] 
  c_R B_{\mu\nu} + {\mathrm {h.c.}} \,, \nn\\
\orru &=& i\, {\overline t}_R \gamma^\mu c_R \Big[H^{\dagger}
  \stackrel{\raisebox{-3pt}{$\leftrightarrow$}}{\raisebox{-.5pt}{$D$}}_{\mu}\! H \Big]  + {\mathrm {h.c.}} \,.
\end{eqnarray}
The brackets mean contraction of $SU(2)$ indices, $Q_3$ and $Q_2$ are the
left-handed $SU(2)$ doublets for the third and second generations, $t_R$ and
$c_R$ are the right-handed $SU(2)$ singlets for the top and charm quarks, $H$ is
the SM Higgs doublet, ${\tilde H} = i \sigma_2 H^*$, and the index $a$ runs over
the $SU(2)$ generators.  The first lower $L$ or $R$ index on the operators
denotes the $SU(2)$ representation of the third generation quark field, while
the second lower index refers to the representation of the first or second
generation field.  In this basis all of the derivatives act on the Higgs
fields.  We could also consider operators directly involving gluons, but
since the indirect constraints on gluonic currents are very weak (see,
e.g.,~\cite{Ferreira:2006jt}), we restrict our focus to the electroweak
operators in  Eq.~(\ref{operators}).   The form of the operators in Eq.~(\ref{operators}) after electroweak symmetry breaking are given in the Appendix.

Throughout the paper we focus on those new operators that contribute to
$t\rightarrow cZ,\,c\gamma$.  In any particular model there may be additional
contributions to Eq.~(\ref{opnorm}) that contribute to $\Delta F=1$ and $\Delta
F=2$ processes in the down sector (e.g., four-fermion operators).  These
operators have suppressed contributions to top FCNCs.  When we bound the
coefficients of the operators in Eq.~(\ref{operators}) from $B$ physics, we
neglect these other contributions.  In any particular model these two sets of
operators may have related coefficients. Unless there are cancellations between
the different operators, the bounds will not get significantly weaker.

There are other dimension-six operators that can mediate FCNC top decays (for
example ${\overline t}_R \gamma^\mu D^\nu c_R B_{\mu\nu}$).  But these can
always be reduced to a linear combination of the operators included in
Eq.~(\ref{operators}) plus additional four-fermion operators and operators
involving $Q_Lq_RHHH$ fields.  For instance, operators involving two quark
fields and three covariant derivatives can be written in terms of operators
involving fewer derivatives using the equations of motion.  Operators involving
two quark fields and two covariant derivatives (e.g., ${\overline Q}_3 D_\mu c_R
D^\mu {\tilde H}$) can be written in terms of operators involving the commutator
of derivatives included in Eq.~(\ref{operators}) plus operators with one
derivative and four-fermion operators.  Finally, operators involving two quark
fields and one covariant derivative can be written in a way that the derivative
acts on the $H$ field, as in Eq.~(\ref{operators}), plus four-fermion operators.  

Of the four-fermion operators which appear after the reduction of the operator
basis, some are suppressed by small Yukawa couplings and can simply be
neglected.  However, some are not suppressed, and of those, the biggest concern
would be semileptonic four-fermion operators, like $(\ov t c)(\ov\ell\ell)$.
These contribute to the same final state as $t\to cZ\to c\ell^+\ell^-$. (We
emphasize $Z \to \ell^+\ell^-$, because the LHC is expected to have the best
sensitivity in this channel~\cite{Carvalho:2007yi,cmstdr}.)  However, the invariant
mass of the $\ell^+\ell^-$ pair coming from a four-fermion operator will have a
smooth distribution and not peak around $m_Z$, so the $Z$-mediated contribution
can be disentangled experimentally. Operators with $(\ov t c)(\ov q q)$ flavor
structure also contribute to $t\to c\ell^+\ell^-$ or $t\to c\gamma$ at one loop,
but their contributions are suppressed by $\alpha/(4\pi)$.  Finally, operators
with the $Q_Lq_RHHH$ structure either renormalize Yukawa couplings, or
contribute to FCNCs involving the Higgs (e.g., $t\to c h$), but we do not
consider such processes, as explained later.

Throughout most of this paper we consider each of the operators one at a time
and constrain its coefficient.  This is reasonable as the operators do not mix under renormalization. One exception is that $\ollu$ and $\ollh$ mix with one another
between the scales $\Lambda$ and $v$, so it would be unnatural to treat them
independently.  Their mixing is given by
\begin{equation}
\frac{\mathrm{d}}{\mathrm{d} \ln \mu} 
  \pmatrix{\cllu(\mu) \cr \cllh(\mu)}
= \frac{3 \alpha_2}{8\pi} 
  \pmatrix{5 & 0 \cr  -4 & 1}
  \pmatrix{\cllu(\mu) \cr \cllh(\mu)} ,
\end{equation}
where $\alpha_2 = \alpha/\sin^2\theta_W$ is the $SU(2)$ coupling.
(The zero in the anomalous dimension matrix is due to the fact that custodial
$SU(2)$ preserving operator $\ollh$ cannot mix into the custodial $SU(2)$
violating $\ollu$.) So, we will also carry out a combined analysis for these two
operators.

We have written the operators in Eq.~(\ref{operators}) in terms of a single SM
Higgs doublet.  In principle there may be many new Higgs scalars, but only those
that acquire a vev will contribute to $t\to cZ$ and $c\gamma$.  Since a triplet
Higgs vev is tightly constrained by electroweak precision tests, we concentrate
on the possibility of multiple Higgs doublets.  With the introduction of extra
Higgs doublets, there are more operators of each particular type ($\ollu$,
$\ollh$, etc.), one linear combination of which gives rise to $t\to c Z$ and
$c\gamma$.  There are also several physical Higgs states that can contribute in
loops in low energy processes.  For each type of operator, a different linear
combination of couplings enter in low energy measurements.  However, without
cancellations this will only differ from the one Higgs case by a number of order
one.  This allows our results to be applied to the general case of multiple
Higgs doublets.\footnote{One possible exception is if an extended Higgs sector
allows Yukawa couplings larger than in the SM, for example, in a two Higgs
doublet model at large $\tan \beta$. Then a Higgs loop may give additional
unsuppressed contributions when we match to the Wilson coefficients at the
electroweak scale.}  Of course, the Higgs sector is also relevant to FCNCs
involving the Higgs, such as $t\to c h$, but we do not consider such processes
as they are more model dependent.

Once we go beyond models with minimal flavor violation (MFV)~\cite{MFV}, the
possibility of new $CP$ violating phases in the NP should be considered.  In MFV
models, top FCNC is not observable at the LHC.  In models such as
next-to-minimal flavor violation (NMFV)~\cite{NMFV} top FCNCs could be
observable and the Wilson coefficients can be complex.  It is not always the
case that the constraints are weaker when the NP Wilson coefficients are real
(in the basis where the up type Yukawa matrix is real and diagonal). Rather,
interference patterns realized in some of the observables mean the constraints
are weakest when some of the new phases are different from $0$ or $\pi$.  We
shall point out the places where phases associated with the new operators can
play an important role and how we treat them. 

In addition to the $B$ physics related constraints we will derive in this paper,
one can also use constraints from electroweak precision observables. However,
these bound flavor-diagonal operators strongly, and the flavor non-diagonal
operators in Eq.~(\ref{operators}) which contribute to top FCNCs are far less
constrained. For instance, the $\ollu$ operator corrects the $W$ propagator at
one loop and so contributes to the $T$ parameter.  The loops involve a $t$ or
$c$ quark, and have one insertion of $\ollu$ and one insertion of $V_{ts}$ or
$V_{cb}$.  Thus, the contribution is suppressed by $|V_{ts}| \sim |V_{cb}| \sim
0.04 $ relative to an insertion of the flavor diagonal equivalent of $\ollu$,
${\overline Q}_3 {\tilde H} D\!\!\!\!\slash {\tilde H}^\dagger  Q_3$.  In
contrast, when considering low energy FCNC processes, $\ollu$ will be more
strongly constrained then its flavor diagonal version.  That is, flavor diagonal
operators are more tightly constrained by electroweak observables than by low
energy FCNCs, while the off diagonal operators are more tightly constrained by
low energy FCNCs.  Moreover, the mixing between these two classes of operators
is small. It occurs at one loop proportional to $y_b^2 |V_{cb}|$, where the
factor of $y_b$,  the bottom Yukawa coupling, is due to a GIM mechanism.  Thus,
we can think of the flavor diagonal and off diagonal operators as independent.
And so for the purpose of studying top FCNCs, we are justified in neglecting
flavor diagonal operators and the relatively weak constraints from electroweak
precision tests.

\section{Weak scale matching}

In this section we derive how the NP operators modify flavor changing
interactions at the electroweak scale and derive  the effective Hamiltonian in
which the $t$, $W$, and $Z$ are integrated out. For numerical calculations we
use besides the Higgs vev, $v=174.1 \gev$, and other standard PDG
values~\cite{pdg}, $|V_{ts}| = 41.0 \times 10^{-3}$~\cite{ckmfitter} and $m_t =
171\gev$~\cite{topmass}.

\subsection{Top quark decays}

After electroweak symmetry is broken, the operators in Eq.~(\ref{operators})
give rise to $t\to cZ$ and $t\to c\gamma$ FCNC decays. The analytic expressions
for the partial widths of these decays are given in Eq.~(\ref{rates}) in the
Appendix.  Numerically, the $t\to c Z$ branching ratio in terms of the Wilson
coefficients is
\begin{eqnarray}\label{eqn:brtcz}
{{\cal B}}(t\to c Z) 
&=& \left(\frac{\lnorm}{\Lambda} \right)^4 \times 10^{-4} \times  
  \Big\{ 1.4 \left[|\clrb|^2+|\crlb|^2 \right]
  - 9.6\, \mathrm{Re}\left( \clrb {\clrw}^{\!\!\!*} +\crlb {\crlw}^{\!\!\!*} \right) \nn\\*
&&{} + 16\left[ |\clrw|^2+|\crlw|^2 \right]
  - 8.3\, \mathrm{Re}\left[ (\cllh+\cllu ){\crlb}^{\!\!\!\!\!*} - \clrb {\crru}^{\!\!\!\!\!*} \right] \nn\\*
&&{} + 28\,\mathrm{Re}\left[ \left(\cllh+\cllu\right) {\crlw}^{\!\!\!*}- \clrw {\crru}^{\!\!\!\!*} \right]
  + 17 \left[ \left|\cllh+\cllu\right|^2+|\crru|^2\right] \Big\} \,.
\end{eqnarray}
The $t c\gamma$ vertex, which has a magnetic dipole structure as required by
gauge invariance, is induced only by the left-right operators. The branching
ratio for $t\to c\gamma$ is
\begin{eqnarray}\label{eqn:brtcgamma}
{\cal B}(t\to c \gamma) = \left( \frac{\lnorm}{\Lambda} \right)^4  
  \times 10^{-4} \times 8.2 \left[ \left|C_{LR}^b+C_{LR}^w\right|^2
  +\left|C_{RL}^b+C_{RL}^w\right|^2 \right] .
\end{eqnarray}
The analogous expressions for $t\to u$ decays are obtained by replacing $C_i$ by
$C'_i$ in Eqs.~(\ref{eqn:brtcz}) and (\ref{eqn:brtcgamma}).

The LHC will have unparalleled sensitivity to such decays.  With $100\,{\rm
fb}^{-1}$ data, the LHC will be sensitive (at 95\% CL) to branching ratios of
$5.5\times 10^{-5}$ in the $t\to cZ$ channel and $1.2\times 10^{-5}$ in the
$t\to c\gamma$ channel~\cite{Carvalho:2007yi}.  In the SM, ${\cal B}(t\to cZ,c\gamma)$
are of order $\alpha (V_{cb} \alpha\, m_b^2/m_W^2)^2 \sim 10^{-13}$, so an
experimental observation would be a clear sign of new physics. 
Equations~(\ref{eqn:brtcz}) and (\ref{eqn:brtcgamma}) will allow one to
translate the measurements or upper bounds on these branching ratios to the
scale of the individual operators.

\subsection{\boldmath $B$ decays}

Many of the operators in Eq.~(\ref{operators}) modify SM interactions
at tree level (this possibility was discussed in
\cite{Buchmuller:1985jz}). After electroweak symmetry breaking,
$\ollu$ gives rise to a ${\overline b} W c$ vertex with the same Dirac
structure as the SM, so the measured value of $V_{cb}$ (which we
denote $V_{cb}^{\rm exp}$) will be the sum of the two.  This allows us
to absorb the new physics contribution of $\cllu$ into the known value
of $V_{cb}^{\rm exp}$ --- in processes where $V_{cb}$ and $\cllu$ enter
the same way, the dependence on $\cllu$ cannot be disentangled.  For
example, the SM unitarity condition, $V_{tb}^*V_{td} + V_{cb}^*V_{cd}
+ V_{ub}^*V_{ud} = 0$, would be violated if one simply shifted the SM
values by the NP contributions. However, the CKM fits have unitarity
built in, so the NP contribution to $V_{cb}$ causes a shift in the values of $V_{ts}$ and $V_{td}$ extracted from the CKM fit, $V_{ts}^{\rm fit}$ and
$V_{td}^{\rm fit}$.  Since we cannot measure all CKM elements independently, we
have to replace $V_{ts}$ and $V_{td}$ by $V_{ts}^{\rm fit}$ and
$V_{td}^{\rm fit}$, plus modified NP contributions.  (Recall that
$V_{ts}$ and $V_{td}$ are only constrained from loop processes where
they enter together with new physics contributions.)  With these
redefinitions we can use $V_{cb}^{\rm exp}$, $V_{ts}^{\rm fit}$ and
$V_{td}^{\rm fit}$ in the CKM fit, and the NP will only have
distinguishable effects in SM loop processes. An analogous procedure
applies to the $t\to u$ contribution to $V_{ub}^{\rm exp}$,
$V_{td}^{\rm fit}$ and $V_{ts}^{\rm fit}$.
Some other operators such as $\clrw$
do not generate a ${\overline b} W c$ vertex with the same Dirac
structure as the SM. Thus, their
contributions to observables from which $V_{cb}$  is extracted may be
disentangled as discussed in the following.

At leading order in the Wolfenstein parameter (Cabibbo angle), $\lambda$, these relations are:
\begin{eqnarray}
V_{cb} &=& V_{cb}^{\rm exp} + (v^2/\Lambda^2)\, \cllu V_{tb}\,, \nn\\
V_{ub} &=& V_{ub}^{\rm exp} + (v^2/\Lambda^2)\, \cllup V_{tb}\,, \nn\\
V_{ts}^* &=& V_{ts}^{* \rm fit}
  - (v^2/\Lambda^2)\, (\cllu V_{cs}^* + \cllup V_{us}^*)\,, \nn\\
V_{td}^* &=& V_{td}^{* \rm fit}
  - (v^2/\Lambda^2)\, (\cllu V_{cd}^* + \cllup V_{ud}^*)\,.
\end{eqnarray}
The $\olrw$ ($\olrwp$) also modifies the $\ov b W c$ ($\ov b W u$) vertex, but
with different Dirac structure from the SM, so its effects can be separated from
the SM contribution.  Finally, $\ollh$ ($\ollhp$) gives tree-level FCNC, since
it contains a ${\overline b} Z s$ ($\ov b Z d$) interaction. 

\begin{figure}[b]
\includegraphics[width=0.32\textwidth]{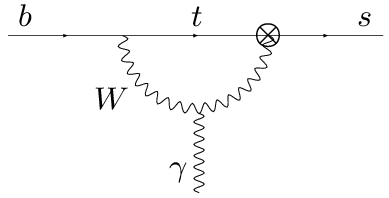}
\caption{A one-loop contribution from $\orlw$ (denoted by $\otimes$) to
$O_{7\gamma}$.}
\label{fig:sketch}
\end{figure}

At the one-loop level, the operators in Eq.~(\ref{operators}) contribute to
$b\to s$ transitions. The constraints from $B$ physics are easiest to analyze by
matching these operators onto operators containing only the light SM fields at a
scale $\mu \sim m_W$.  We use the standard basis as defined in~\cite{BBL}.  
Integrating out the top, $W$, and $Z$, the most important operators for $B\to
X_s \gamma$ and $B\to X_s \ell^+ \ell^-$ which are affected by NP are
\begin{eqnarray}
O_{7\gamma} &=& \frac{e}{8\pi^2}\,
  [m_b {\overline s}\, \sigma^{\mu\nu} (1+\gamma_5) b]\, F_{\mu\nu}\,, \nn\\*
O_{9V} &=& [{\overline s}\gamma^\mu (1-\gamma_5)b]\, 
  [{\overline \ell} \gamma_\mu \ell]\,, \nn\\*
O_{10A} &=& [{\overline s}\gamma^\mu (1-\gamma_5)b]\, 
  [{\overline \ell} \gamma_\mu \gamma_5 \ell]\,.
\end{eqnarray}
For example, the diagram in Fig.~\ref{fig:sketch} gives a contribution
from $\orlw$ (denoted by $\otimes$) to $O_{7\gamma}$.  The coefficients of the
QCD and electroweak penguin operators, $O_{3,\ldots,10}$, are also modified, but
their effect on the processes we consider are suppressed.

Summing the relevant diagrams, the contributions of all operators can be
expressed in terms of generalized Inami-Lim functions, presented in the
Appendix.  Setting $\Lambda=1\tev$, the numerical results
are\footnote{Throughout this paper we will bound $C_i (\lnorm/\Lambda)^2$ and
quote numerical results setting $\Lambda = \lnorm$.} 
\begin{eqnarray}
C_{7\gamma}(m_W) &=& - 0.193 + \left( 0.810 \cllu + 0.179 \cllh 
  + 0.310 \crlw - 0.236 \crlb + 0.004 \clrw - 0.003 \clrb \right) , \nn\\*
C_{9V}(m_W) &=& \frac{\alpha}{2\pi} \left[ 1.56
  + \left( -0.562 \cllu + 44.95 \cllh 
  - 0.885 \crlw - 1.127 \crlb + 0.046 \clrw + 0.004 \clrb \right)\right] ,\nn\\*
C_{10A}(m_W) &=& \frac{\alpha}{2 \pi} \left[ - 4.41  
  + \left( - 7.157 \cllu - 598 \cllh + 3.50 \crlw - 0.004 \crru \right)\right] .
\end{eqnarray}
The first term in each expression is the SM contribution. Note that the $\ollh$
contribution is large because it is at tree level, while $\olrb$, $\olrw$, and
$\orru$ are tiny because they are suppressed by $m_c / m_W$ and so the
constraints on these will be weaker.  In the case of $b \to d$ transitions the
NP contribution has to be rescaled by the ${\cal O}(1/\lambda)$ factor,
$|V_{ts}^* V_{ud} / V_{td}^* V_{cs}| \approx 5.6$, and $C_i$ should be replaced with $C_i^\prime$.

\subsection{\boldmath $\Delta F=2$ transitions}
\label{sec:df2}

The operators $\ollu$, $\cllh$, and  $\orlw$ also contribute to $\Delta F=2$
transitions, i.e., neutral meson mixings. Again, the contribution from $\ollh$
is present at tree level, while the other two contribute starting at one-loop
order. The relevant functions are again listed in the Appendix. The
modifications relative to the SM Inami-Lim function can be parameterized as
$S_0\to S_0(1+h_M e^{2i\, \sigma_M})$ for each neutral meson system.
Numerically (setting $\Lambda=\lnorm$), for $B_s^0\ov B{}_s^0$ mixing, the effect of the $t\to c$ operators
is given by
\begin{equation}\label{eq:df2}
h_{B_s} e^{2i \sigma_{\!B_{\!s}}} = 800 (\cllh)^2 + 0.92 \cllh \cllu 
   - 6.84 (\cllu)^2  + 1.55 \cllh - 2.64 \cllu - 0.32 (\crlw)^2 - 1.03 \crlw\,.
\end{equation}
The contributions of the $O'_i$ operators to $B_s^0\ov B{}_s^0$ mixing is given
by replacing $C_i$ with $C'_i$ in Eq.~(\ref{eq:df2}) and multiplying its right-hand
side by $\lambda$.

The contribution of the $O_i$ operators to $B_d^0\ov B{}_d^0$ mixing  is
obtained by multiplying the right-hand side of Eq.~(\ref{eq:df2}) by $e^{i
\beta}$, where $\beta$ is the CKM phase, $\beta = \arg(-{V_{cd}V_{cb}^*/
V_{td}V_{tb}^*})$.  Whereas the contribution of the $O'_i$ operators to
$B_d^0\ov B{}_d^0$ mixing is obtained again by replacing $C_i$ with $C'_i$ in
Eq.~(\ref{eq:df2}) and multiplying its right-hand side by $-e^{i\beta}/\lambda$.

Finally, the $O'_i$ contribution to $K^0\ov K{}^0$ mixing is the same as that
to $B_d^0\ov B{}_d^0$ mixing, up to corrections suppressed by powers of
$\lambda$.  For the $O_i$ contribution to $K^0\ov K{}^0$ mixing, one has to
replace in Eq.~(\ref{eq:df2}) each Wilson coefficient $C_i$ by $C_i+C_i^*
e^{i\beta}$ (see Eq.~(\ref{kappas}) in the Appendix), and add to it the
additional contribution
\begin{equation}
\Delta (h_K e^{2 i \sigma_K}) = 2.26\, {\rm Re}({\cllh}\cllu)\, e^{i\beta}
  - 5.17\, |\cllu|^2\, e^{i \beta} - 8.35\, |\crlw|^2\, e^{i\beta}\,.
\end{equation}
These expressions are valid up to corrections suppressed by $\lambda^2$ or more.

\section{Experimental Constraints}
\label{down}

In this section we use low energy measurements to constrain the Wilson
coefficients of the operators in Eq.~(\ref{operators}).  Throughout we assume
that there are no cancellations between the contributions from different
operators.

\subsection{Direct bounds}

The best direct bounds on the operators in Eq.~(\ref{operators}), as summarized
in~\cite{pdg}, come at present from searches for FCNCs at the Tevatron, LEP, and
HERA. The strongest direct constraints on $t\to cZ$ and $t\to uZ$ come from an
OPAL search for $e^+ e^-\to \ov t c$ in LEP~II~\cite{Abbiendi:2001wk}. The upper
limit on the branching ratio ${\cal B}(t\to cZ,\, u Z)<0.137$ bounds the $LL$
and $RR$ operators. For neutral currents involving a photon, there is a
constraint from ZEUS that looked for $e^\pm p\to e^\pm t
X$~\cite{Chekanov:2003yt}. This bounds ${\cal B}(t\to u \gamma) < 0.0059$, and
is the strongest constraint on the $RL$ and $LR$ operators with an up quark. 
The other bounds come from a CDF search in Tevatron Run~I, which bounds ${\cal
B}(t\to c\gamma,\, u\gamma) < 0.032$~\cite{Abe:1997fz} and constrains the $LR$
and $RL$ involving a charm.  We translate these branching ratios into bounds on
the Wilson coefficients and list them in the first rows of
Tables~\ref{tab:Cconstraints} and \ref{tab:Uconstraints}. The LHC reach with
100\,fb$^{-1}$ data, as estimated in the ATLAS study~\cite{Carvalho:2007yi} is ${\cal
B}(t\to cZ,\, uZ) < 5.5 \times 10^{-5}$ and ${\cal B}(t\to c\gamma,\, u\gamma) <
1.2 \times 10^{-5}$. These will improve the current direct constraints on the
Wilson coefficients by one and a half orders of magnitude, as summarized in the
second rows of the tables.

\subsection{\boldmath $B\to X_s\gamma$ and $B\to X_s \ell^+\ell^-$}
\label{sec:bsgll}

We first consider the constraints from $B\to X_s\gamma$. At the scale $m_b$,
$O_{7\gamma}$ gives the leading contribution. Using the NLO SM formulae from
Ref.~\cite{bsgrange}, we obtain
\begin{equation}
{\cal B}(B\to X_s \gamma) = 10^{-4} \times 
  \left( 0.07 + \big| 1.807 + 0.081\, i 
  + 1.81\, \Delta C_{7\gamma}(m_W) \big|^2 \right) ,
\end{equation}
where $\Delta C_{7\gamma}(m_W)$ is the NP contribution to $C_{7\gamma}$ at the
$\mu=m_W$ matching scale. The current experimental average~\cite{hfag}, ${\cal
B}(B\to X_s \gamma) = (3.55 \pm 0.26) \times 10^{-4}$, implies at 95\%
CL\footnote{Hereafter all constraints are quoted at 95\% CL, unless otherwise
specified.} (setting $\Lambda = \lnorm$)
\begin{eqnarray}\label{bsgsol1}
-0.07 < \cllu < 0.04   &\qor&   1.2 < \cllu < 1.3\,, \nn\\*
-0.3 < \cllh < 0.16    &\qor&  5.3 < \cllh < 5.8\,, \nn\\*
-0.2 < \crlw < 0.1    &\qor&   3.1 < \crlw < 3.4\,,\nn\\*
-0.1 < \crlb < 0.24   &\qor&  -4.5 < \crlb < -4.1\,,
\end{eqnarray}
The first (left) intervals are consistent with the SM, while the second (right)
ones require new physics at the ${\cal O}(1)$ level.  The non-SM region away
from zero is disfavored by $b\to s \ell^+\ell^-$ discussed below, but we include
it here for completeness. For the operators whose contributions are suppressed
by $m_c$, we find
\begin{eqnarray}\label{bsgsol2}
-14 < \clrw < 7\,, \qquad -10 < \clrb < 19\,,
\end{eqnarray}
and no meaningful bound for $\crru$.  To obtain the results in
Eq.~(\ref{bsgsol1}) and (\ref{bsgsol2}), we assumed that the NP contributions
are real relative to the SM, i.e., that there are no new $CP$ violating phases. 
Had we not made this assumption, the allowed regions would be annuli in the
complex $C_i$ planes.

Next we consider $B\to X_s \ell^+\ell^-$. The theoretically cleanest bound at
present comes from the inclusive $B\to X_s \ell^+\ell^-$ rate measured for
$1\gev^2 < q^2 < 6\gev^2$~\cite{Aubert:2004it}
\begin{equation}
{\cal B}(B\to X_s \ell^+\ell^-)_{1\gev^2<q^2 <6\gev^2}
  = (1.61 \pm 0.51)\times 10^{-6}\,.
\end{equation}
Due to the unusual power counting in $B\to X_s \ell^+\ell^-$, the full set of
${\cal O}(\alpha_s)$ corrections are only included in what is called NNLL order,
achieving an accuracy around 10\%. For the SM prediction we use the NNLL
calculation as implemented in Ref.~\cite{Lee:2006gs}. This calculation does not
normalize the rate to the $B\to X\ell\bar\nu$ rate; doing so would not improve
the prediction significantly and would unnecessarily couple different operators'
contributions.  We include the modifications of $C_{7\gamma}$, $C_{9V}$, and
$C_{10A}$ due to the new operators at lowest order.  With our input parameters,
we obtain
\begin{eqnarray}
{\cal B}(B\to X_s \ell^+ \ell^-)_{1<q^2 <6\gev^2} 
= 10^{-6} &\times& \Big\{ 1.55 
  + 35100\, \big[ |\Delta C_{9V}(m_W)|^2 + |\Delta C_{10A}(m_W)|^2 \big] 
  + 0.45\, |\Delta C_{7\gamma}(m_W)|^2 \nn\\
&&{} + {\rm Re} \big[(180 + 5 i) \Delta C_{9V}(m_W) \big] 
  - 360\, {\rm Re} \big[ \Delta C_{10A}(m_W) \big] \\
&&{} - {\rm Re} \big[(0.17+0.04 i) \Delta C_{7\gamma}(m_W) \big]
  - 200\, {\rm Re} \big[\Delta C_{9V}(m_W)^* \Delta C_{7\gamma}(m_W)\big]
  \Big\}\,. \nn
\end{eqnarray}
The simplest way to proceed would be to bound $C_{7\gamma}$, $C_{9V}$, and
$C_{10A}$ separately at $\mu=m_W$, assuming that the others have their SM
values, and use this to constrain new physics.  This procedure would not be
consistent, since the NP necessarily affects these Wilson coefficients in a
correlated way. Instead, we directly constrain the coefficients of $\ollu$,
$\ollh$, $\olrw$, and $\olrb$, which also yields stronger constraints. With
$\Lambda=\lnorm$, we obtain
\begin{eqnarray}\label{bsllsol}
-1.1 &<& \cllu <  0.3 \,,\nn\\
-1.8 \times 10^{-2} &<& \cllh <  -1 \times 10^{-2} \qor -5 \times 10^{-3} 
  < \cllh < 3 \times 10^{-3} \,,\nn\\
-0.5 &<& \crlw <  0.7 \,\qquad\quad\qor\qquad\quad~~\, 1.7 < \crlw< 3 \,,\nn\\
-2.0  &<& \crlb <  3.5 \,.
\end{eqnarray}

\begin{figure}[t]
\includegraphics[width=0.45\textwidth,clip]{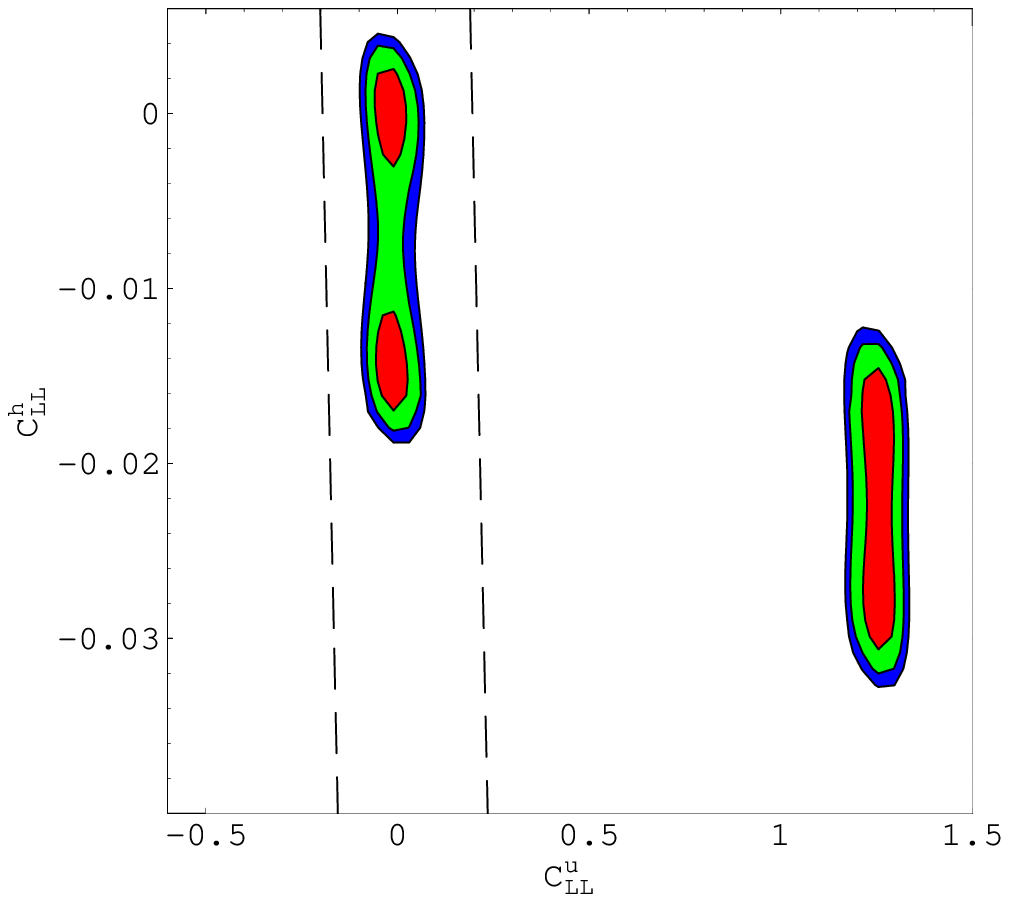}
\caption{Constraints from $B\to X_s \gamma$ and $B\to X_s \ell^+\ell^-$ in the
$\cllu$~--~$\cllh$ plane. The red, green, and blue regions denote 68\%, 95\%,
and 99\% CL, respectively. The region between the dashed lines is beyond the LHC
sensitivity.}
\label{fig:clluhmix}
\end{figure}

The combined constraints from $B\to X_s\gamma$ and $B\to X_s\ell^+\ell^-$ on
these four Wilson coefficients are shown in Table~\ref{tab:Cconstraints} in the
Conclusions.  We plot in Fig.~\ref{fig:clluhmix} the bound on the $LL$
operators in the $\cllu$~--~$\cllh$ plane.  The SM corresponds to the point
(0,\,0). A measurement or a bound on the $t\to c Z$ branching ratio corresponds to a
nearly vertical band. The LHC is sensitive to this whole plane, except for the
band between the dashed lines.

The above bounds were derived assuming that the NP contribution is real relative
to the SM.  It is conceivable that improved measurements of $B\to
X_s\ell^+\ell^-$ will lead to constraints on the $CP$ violating phases before
the LHC is be able to probe top FCNCs.  Thus we postpone a full analysis with
complex NP Wilson coefficients until more data become available.

\subsection{\boldmath Exclusive and inclusive $b\to c\ell\bar\nu$ decays}

In this subsection we investigate the constraints on the operators in
Eq.~(\ref{operators}) due to measurements of semileptonic $b\to c$ decays. They
will allow us to bound the coefficient of the operator $\olrw$, which contains a
right handed charm field and is weakly constrained otherwise. We focus on three
types of constraints coming from the ratio of exclusive $D$ and $D^*$ rates, the
polarization in the $D^*$ mode, and moments in inclusive spectra.

We begin with the exclusive case where the $B\to D \ell\nu$ and $B\to D^*
\ell\nu$ rates can be calculated in an expansion in $\lqcd/m_{b,c}$ using heavy
quark effective theory.  The form factors at zero recoil, where $w=v\cdot
v' =1$ ($v$ and $v'$ are the four-velocities of the $B$ and $D^{(*)}$ mesons,
respectively), have been determined from lattice QCD~\cite{bclqcd}.  In the SM
the ratio of rates is independent of $V_{cb}$, and therefore it provides a
good test for non-SM contributions. The presence of the new operator, $\olrw$,
affects the two rates differently. The rates are given by~\cite{manowise}
\begin{eqnarray}\label{BDrate}
{\d\Gamma ({B\to D\ell\nu})\over \d w} &=& \frac{G_F^2 m_B^5}{48\pi^3}\, r^3 
  (w^2-1)^{3/2} (1+r)^2 |V_{cb}|^2 ({\cal F}_D)^2 \,, \nn\\*
{\d\Gamma({B\to D^*\ell\nu})\over \d w} &=& {G_F^2 m_B^5\over 48 \pi^3}\, r^{*3}
  \sqrt{w^2-1}\, (1+w)^2 \left[(1-r^*)^2 + \frac{4w}{1+w}\, 
  (1-2wr^*+r^{*2}) \right] |V_{cb}|^2 ({{\cal F}_{D^*}})^2\,,
\end{eqnarray}    
where $r=m_D/m_B$ and $r^*=m_{D^*}/m_B$. The form factors ${\cal F_D}$ and
${\cal F_{D^*}}$ can be decomposed in terms of 6 form factors,
$h_{+,-,V,A_1,A_2,A_3}$~\cite{manowise}.  At leading order in the heavy quark
limit ${\cal F}(w) = {\cal F}^*(w) = h_{+,V,A_1,A_3} = \xi(w)$, where $\xi(w)$
is the Isgur-Wise function~\cite{IW}, while $h_{-,A_2} = 0$.  Therefore,
it is useful to define the following ratios of form factors
\begin{equation}\label{R12def}
R_1(w) = \frac{h_V}{h_{A_1}} \,, \qquad 
  R_2(w) =\frac{h_{A_3}+r^* h_{A_2}}{h_{A_1}} \,,
\end{equation}
which are equal to unity in the heavy quark limit and have been measured
experimentally.  

Following the analysis of~\cite{Walter}, we can absorb the new physics
contributions in the form factors.  We obtain
\begin{equation}\label{BDffNP}
\begin{array}{rclrcl}
\Delta h_+ &=& k (1+r) (1-w) \xi(w)\,, \qquad &
  \Delta h_- &=& - k (1-r) (1+w) \xi(w)\,, \\
\Delta h_{A_1} &=&  -2 k (1-r^*) \xi(w)\,, &
  \Delta h_{A_2} &=&  -2k \xi(w)\,, \\
\Delta h_{A_3} &=&  -2k \xi(w)\,, &
  \Delta h_V &=& 2k(1+r^*) \xi(w)\,,
\end{array}
\end{equation}
where
\begin{equation}
k = \frac{2\, v\, m_B}{\Lambda^2}\, 
  {\rm Re}\bigg(\frac{\clrw V_{tb}}{V_{cb}}\bigg) \,.
\end{equation}
For the new physics contribution we include only the leading term, so we set
$\xi(1)=1$. Setting $\Lambda = \lnorm$, we obtain
\begin{eqnarray}
{\cal F}_D(1) 
&\approx &{\cal F}^{\rm SM}_D(1) - 1.01 \times 10^{-3} \times
  {\rm Re}\bigg(\frac{\clrw V_{tb}}{V_{cb}}\bigg) \,, \nn\\*
{\cal F}_{D^*}(1) &\approx & {\cal F}^{\rm SM}_{D^*}(1) - 2.02 \times 10^{-3}
  \times {\rm Re}\bigg(\frac{\clrw V_{tb}}{V_{cb}}\bigg) \,, \nn\\
R_1(1) &\approx & R_1^{\rm SM}(1) + 6.52 \times 10^{-3} \times
  {\rm Re}\bigg(\frac{\clrw V_{tb}}{V_{cb}}\bigg) \,, \nn\\*
R_2(1) &\approx & R_2^{\rm SM}(1) - 2.48 \times 10^{-3} \times
  {\rm Re}\bigg(\frac{\clrw V_{tb}}{V_{cb}}\bigg) \,.
\end{eqnarray}

Recent lattice QCD calculations~\cite{bclqcd} give ${\cal F}^{\rm SM}_D(1) =
1.074\pm 0.024$ and ${\cal F}_{D^*}^{\rm SM}(1)=0.91\pm0.04$. For $R_1^{\rm SM}$
and $R_2^{\rm SM}$ we use the results of~\cite{Grinstein:2001yg}, scanning over
the hadronic parameters that enter. The experimental results are $|V_{cb}| {\cal
F}_D(1) = (42.4\pm4.4)\times 10^{-3}$, $|V_{cb}| {\cal F}_{D^*}(1) =
(36.2\pm0.6)\times 10^{-3}$~\cite{hfag}, $R_1(1) = (1.417\pm0.075)$, and $R_2(1)
=(0.836\pm0.043)$~\cite{Aubert:2006mb}. We set $|V_{tb}|=1$ and do a combined
fit for $\clrw$ and $|V_{cb}|$. We find
\begin{equation}
-0.2 < \frac{{\rm Re}(V_{cb}^*\, \clrw V_{tb})}{|V_{cb}|} < 1.6\,.
\end{equation}

We next turn to inclusive $B\to X_c\ell\bar\nu$ decays, which is also sensitive
to the presence of the additional operators. We concentrate on the partial
branching ratio and moments constructed from the charged lepton energy spectrum
(see, e.g.,~\cite{Gremm:1996yn}), 
\begin{equation}
M_0(E_0) = \tau_B \int_{E_0} {\d\Gamma\over \d E_\ell}\, \d E_\ell \,, 
  \qquad
M_1(E_0) = {\int_{E_0} E_\ell {\d\Gamma\over \d E_\ell}\, \d E_\ell 
  \over \int_{E_0} {\d\Gamma\over \d E_\ell}\, \d E_\ell} \,, \qquad
M_k(E_0) = {\int_{E_0} [E_\ell - M_1(E_0)]^k {\d\Gamma\over \d E_\ell}\, \d E_\ell 
  \over \int_{E_0} {\d\Gamma\over \d E_\ell}\, \d E_\ell} \,.
\end{equation}
These are well measured and can be reliably calculated.  We use the SM
prediction including $1/m_b^2$ and $\alpha_s$ corrections and compare it in a
combined fit with the 20 Babar~\cite{Aubert:2004td} and a subset~\cite{phillip}
of the 45 Belle~\cite{Urquijo:2006wd} measurements, including their
correlations.  The modification of $\d\Gamma/ \d E_\ell$ due to the $\clrw$
coupling is given by
\begin{eqnarray}\label{bcinclNP}
{\d\Gamma^{\rm NP}(B\to X_c\ell\bar\nu)\over \d y} =
&-& \frac{G_F^{5/2} m_b^6\, v^2\, {\rm Re}(\clrw V_{cb})}{6 \sqrt[4]2\, \pi^3 
 \Lambda^2}\, \frac{\sqrt{\rho}\, y^2 (3-y) (1-y-\rho)^3}{(1-y)^3} \nn\\*
&+& \frac{\sqrt2 G_F^3 m_b^7\, v^4\, |\clrw|^2}{3 \pi^3 \Lambda^4}\,
  \frac{y^2 (3-y) (1-y-\rho)^4}{(1-y)^3}\,,
\end{eqnarray}
where $y=2E_\ell/m_b$ and $\rho=m_c^2/m_b^2$.  It is known that the data cannot
be fitted well with the OPE truncated at $1/m_b^2$.  Including the $1/m_b^3$
corrections in a more complicated fit would make the agreement with the SM
better, and therefore our bounds stronger.

\begin{figure}[t]
\includegraphics[width=0.45\textwidth,clip]{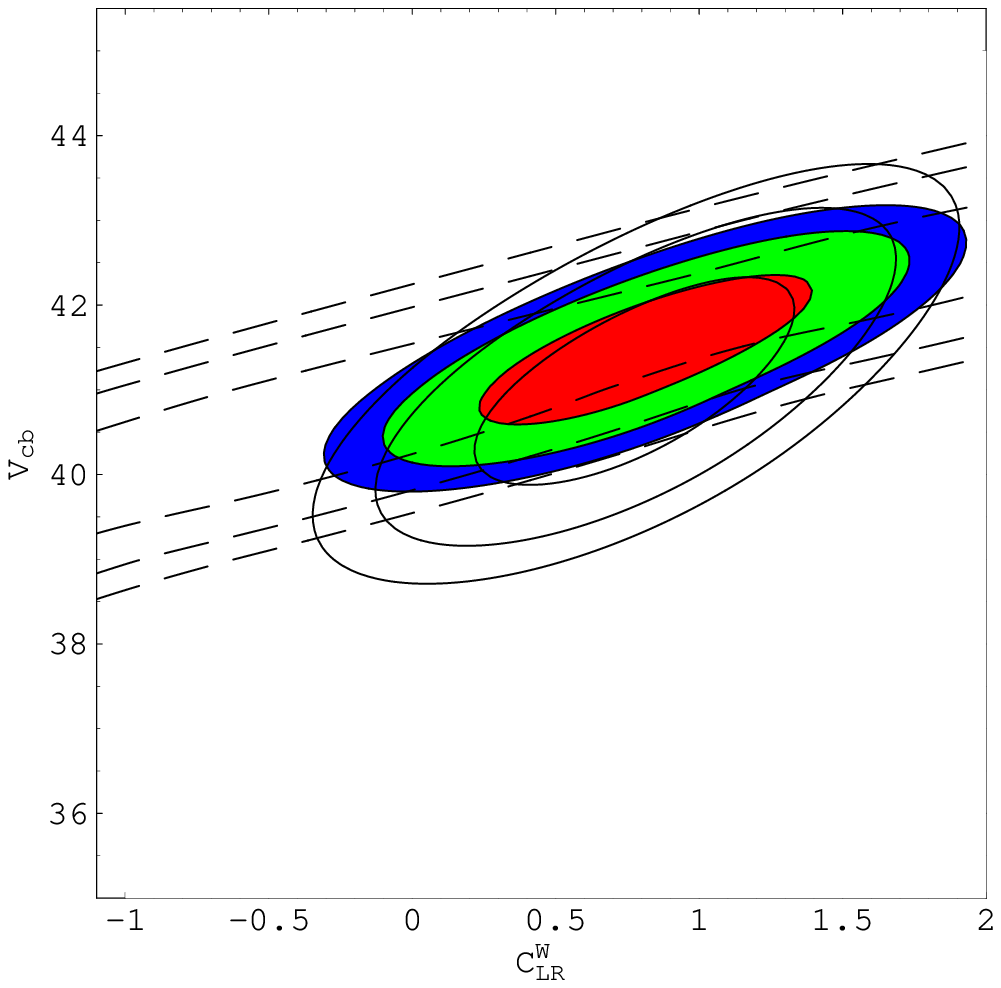}
\caption{Constraints on $\olrw$ in the ${\rm Re}(\clrw)$~--~$|V_{cb}|$ plane
from semileptonic $B\to X_c \ell\bar\nu$ (solid curves) and $B\to
D^{(*)}\ell\bar\nu$ decays (dashed curves) and their combination (shaded
areas).  For each constraint the 68\%, 95\% and 99\% CL regions are shown.}
\label{fig:btoc-excl}
\end{figure}

The combined constraints on $\clrw$ and $|V_{cb}|$ from exclusive and inclusive
decays is shown in Fig.~\ref{fig:btoc-excl}. The solid curves show the
constraints from inclusive decays, the dashed curves show the bounds from
exclusive semileptonic decays to $D$ and $D^*$, and the shaded regions show the
combined constraints (the confidence levels are as in Fig.~\ref{fig:clluhmix}).

\subsection{\boldmath Exclusive and inclusive $b\to u\ell\bar\nu$ decays}

We now turn to some 3rd~$\to$~1st generation transitions.  While the
experimental constraints are less precise for these than for 3rd~$\to$~2nd
generation transitions, the SM also predicts smaller rates, and therefore NP
could more effectively compete with the SM processes.  These constraints are
particularly important as they bound the $O'_i$ contributions relevant for $t\to
u$ decays, which might not be distinguishable at the LHC from $t\to c$.

As is the case for 3rd~$\to$~2nd generation transitions, exclusive and
inclusive  semileptonic $b\to u$ decays can constrain the operator $\olrwp$ in
$t\to u$ transition. Similarly to $b\to c\ell\bar\nu$, this operator distorts
the lepton energy spectrum, so information on the lepton energy moments could
constrain it. However, such measurements are not yet available for $B\to X_u
\ell\bar\nu$. Therefore, to distinguish between the SM $V_{ub}$ contribution and
$\clrw$, we use $B\to \pi \ell\bar\nu$ in addition to the inclusive data. 

For exclusive $B\to \pi \ell\bar\nu$ decay, we use for the SM prediction the
parameterization of Ref.~\cite{Arnesen:2005ez}, which relies on analyticity
constraints and lattice QCD calculations of the form factors at large
$q^2$~\cite{Okamoto:2004xg,Shigemitsu:2004ft}. The NP contribution is
\begin{eqnarray}
\frac{\d\Gamma^{\rm NP}(B\to \pi \ell\bar\nu)}{\d q^2} 
&=& \frac{G_F^2|p_\pi\!|^3}{24\pi^3}\, \bigg\{ 
  \frac{4m_B^2 v^2 |\clrwp|^2}{\Lambda^4} \left[ (1+\hat q^2) f_- 
  + (1-\hat q^2) f_+ \right]^2 \nn\\*
&&{}\qquad\qquad - \frac{4m_B v\, {\rm Re}(V_{tb} \clrw V_{ub}^*)}{\Lambda^2}
  \left[ (1-\hat q^2) f_+^2 + (1+\hat q^2) f_- f_+ \right] \bigg\} \,.
\end{eqnarray}
where the $f_\pm$ form factors are functions of the dilepton invariant mass,
$q^2$, $\hat q^2 = q^2/m_B^2$, and we neglected terms suppressed by
$m_\pi^2/m_B^2$.

For inclusive $B\to X_u\ell\bar\nu$ decay, we focus on the measurement utilizing
combined cuts~\cite{bll} on $q^2$ and the hadronic invariant mass, $m_X$, and
compare it with the Belle and Babar measurements~\cite{Kakuno:2003fk}.  Using
this determination of $V_{ub}$ is particularly simple for our purposes, because
in the large $q^2$ region the mild cut on $m_X$ used in the analysis only
modifies the rate at a subleading level.  Working to leading order in the NP
contribution, we can neglect the effect of the $m_X$ cut on the NP and include
the NP contribution to the rate via
\beq
\frac{\d\Gamma^{\rm NP}(B\to X_u \ell\bar\nu)}{\d q^2} 
  = \frac{G_F^2 m_b^5}{192 \pi^3}\, \frac{32 m_B^2 v^2}{\Lambda^4}\, 
  |\clrwp|^2\, \hat q^2 (\hat q^2-1)^2 (\hat q^2+2)\,. 
\eeq
Since the interference between the SM and NP is suppressed by $m_u/m_b$ (see the
$\sqrt\rho$ factor in Eq.~(\ref{bcinclNP}) in the first term), there is no
dependence on the weak phase of $\clrwp$ in the inclusive decay.  Using other
determinations of $V_{ub}$ would be harder to implement and would not change our
results significantly.

\begin{figure}[t]
\includegraphics[width=0.45\textwidth,clip]{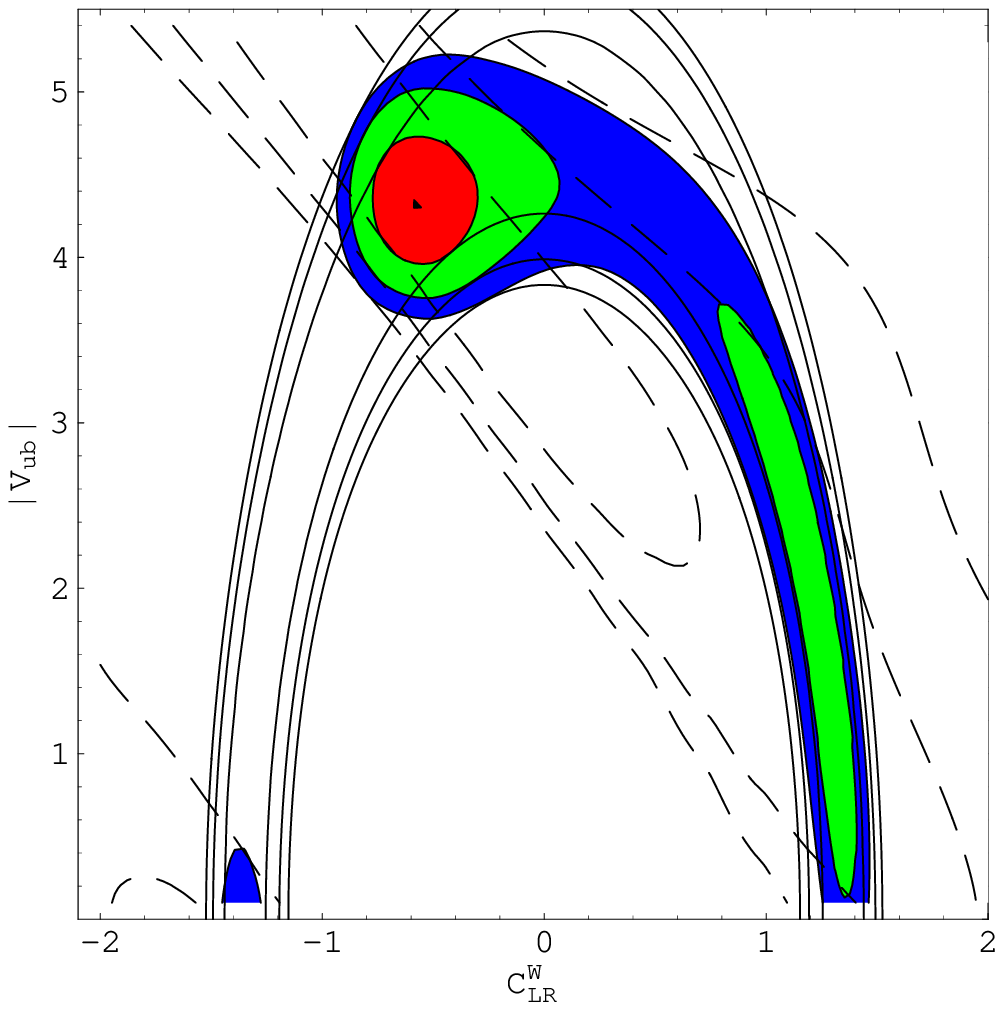}
\caption{Constraints  on $\olrwp$ in the ${\rm Re}(\clrwp)$ -- $|V_{ub}|$ plane
from $B\to X_u \ell\bar\nu$ (solid curves) and $B\to \pi \ell\bar\nu$ (dashed
curves) and their combination (shaded areas). For each constraint the 68\%, 95\%
and 99\% CL regions are shown.}
\label{fig:btou-incexcl}
\end{figure}

The combined constraint on $\clrwp$ and $|V_{ub}|$ from inclusive and exclusive
decays is shown in Fig.~\ref{fig:btou-incexcl}. (This uses the lattice QCD input
from Fermilab~\cite{Okamoto:2004xg}, and the one using the HPQCD
calculation~\cite{Shigemitsu:2004ft} would also be similar.)

\subsection{\boldmath $B\to \rho\gamma$ and $B\to \mu^+\mu^-$}

The inclusive $B\to X_d\gamma$ decay has not been measured yet, and there is
only limited data on $B\to \rho\gamma$. Averaging the
measurements~\cite{Abe:2005rj} using the isospin-inspired\footnote{Isospin is
not a symmetry of the electromagnetic interaction. This average relies on the
heavy quark limit to argue that the dominant isospin violation is $\lqcd/m_b$
suppressed.  With more precise data, using only $B^0$ decays will be
theoretically cleaner, because annihilation is suppressed in the $B^0$ compared
to the $B^\pm$ modes.  At present, this would double the experimental error, so
we include the $B^\pm$ data.} relation ${\cal B}(B\to\rho\gamma) = {\cal
B}(B^\pm\to\rho^\pm\gamma) = 2 (\tau_{B^\pm}/\tau_{B^0})\, {\cal
B}(B^0\to\rho^0\gamma)$, and the PDG value $\tau_{B^\pm}/\tau_{B^0} = 1.07$, we
obtain
\begin{equation}
{\cal B}(B\to\rho\gamma) = (1.26 \pm 0.23) \times 10^{-6}\,. 
\end{equation}
To reduce the sensitivity to form factor models, we normalize this rate to
${\cal B}(B\to K^*\gamma) = \left[ {\cal B}(B^\pm\to K^{*\pm}\gamma) +
(\tau_{B^\pm}/\tau_{B^0})\, {\cal B}(B^0\to K^{*0}\gamma) \right]/2 = (41.4 \pm
1.7) \times 10^{-6}$,
\beq\label{brhogamma}
\frac{{\cal B}(B\to\rho\gamma)}{{\cal B}(B\to K^*\gamma)} = 
  \bigg|{V_{td}\over V_{ts}}\bigg|^2\,
  \bigg(\frac{m_B^2 - m_\rho^2}{m_B^2 - m_{K^*}^2}\bigg)\, \xi_\gamma^{-2}
  \frac{|C_{7\gamma}|^2}{|C_{7\gamma}^{\rm SM}|^2}\,.
\eeq
We use $\xi_\gamma = 1.2 \pm 0.15$, where this error estimate accounts for the
fact that we consider the rates to be determined by $O_{7\gamma}(m_b)$ alone.
The contributions of other operators have larger hadronic uncertainties and are
expected to partially cancel~\cite{Ball:2006eu}. If first principles lattice QCD
calculations of the form factor become available then one can avoid taking the
ratio in Eq.~(\ref{brhogamma}), and directly compare the calculation of ${\cal
B}(B\to\rho\gamma)$ with data. We obtain the following constraints
\begin{eqnarray}\label{bdgsol}
-0.26 < \cllup < -0.21   &\qor&   -0.026 < \cllup < 0.03\,, \nn\\
-1.2 < \cllhp < -0.9    &\qor&  -0.11 < \cllhp < 0.13\,, \nn\\
-0.7 < \crlwp < -0.5    &\qor&   -0.07 < \crlwp < 0.08\,,\nn\\
-0.1 < \crlbp < 0.09   &\qor&  0.7 < \crlbp < 0.9\,. 
\end{eqnarray}
Note that there are no constraints on $\olrwp$ or $\olrbp$, because of their
$m_u/m_W$ suppression. As for $B\to X_s\gamma$, the two solutions in
Eq.~(\ref{bdgsol}) correspond to the sign ambiguity in interpreting the
constraint on $|C_{7\gamma}|^2$ when we assume that the NP contributions are
real relative to the SM.  Had we not made this assumption, the allowed regions
would be annuli in the complex $C'_i$ planes.

The NP operators we consider also contribute to the rare decays $B_{d,s}\to
\mu^+\mu^-$. This is most interesting for $B_d\to \mu^+\mu^-$, since one expects
that the NP contribution is enhanced compared to the SM by
$[(v^2/\Lambda^2)(1/|V_{td}|)]^2$, which is around 20 for $\Lambda = 1\tev$.
Moreover, $\ollhp$ contributes at tree level, so its contribution is enhanced by
an additional factor of $(4\pi/\alpha)^2$. Although this decay mode has not yet
been observed and the present upper bound ${\cal B}(B\to \mu^+\mu^-) < 3 \times
10^{-8}$~\cite{cdfbsmumu} is two orders of magnitude above the SM expectation,
it still gives a useful constraint on $\ollhp$. In particular, for $\Lambda =
\lnorm$, we obtain
\begin{equation}
-0.023 < \cllhp <0.026\,.
\end{equation}

\begin{figure}[t]
\includegraphics[width=0.5\textwidth,clip]{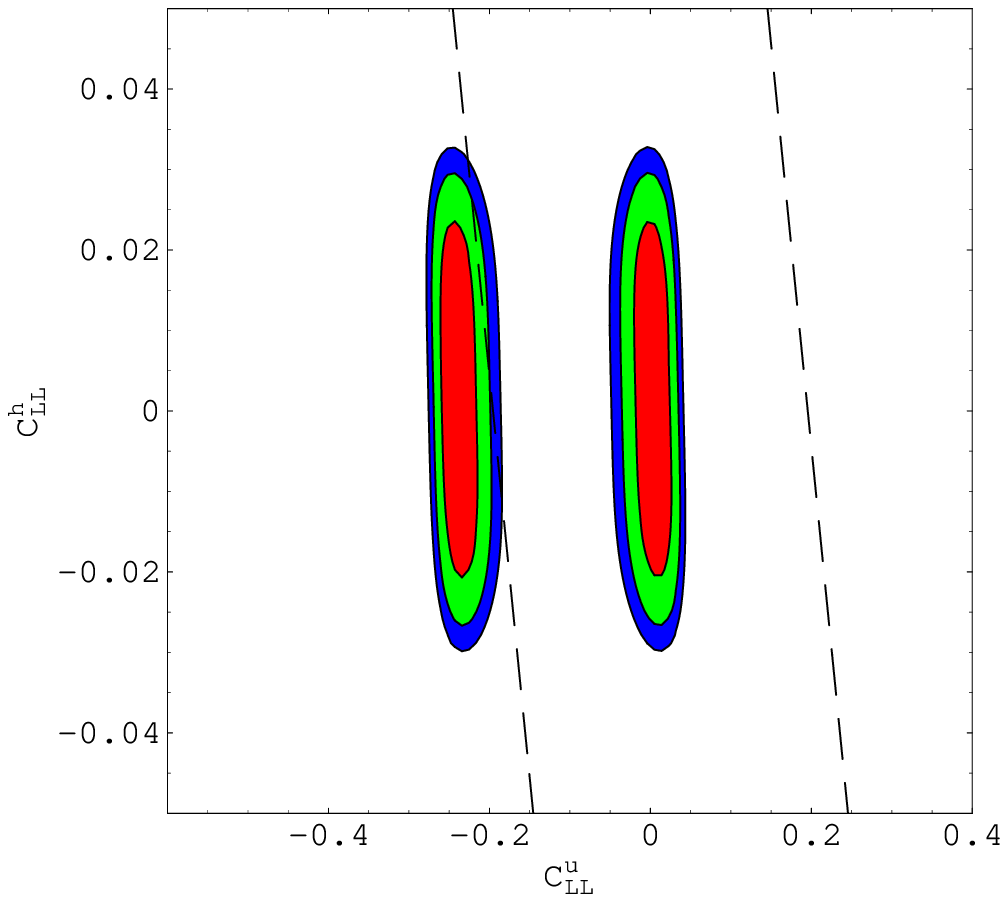}
\caption{Constraints from $B\to \rho \gamma$ and $B \to \mu^+\mu^-$ in the
$\cllup$~--~$\cllhp$ plane. The red, green, and blue regions denote 68\%, 95\%,
and 99\% CL, respectively. The region between the dashed lines is beyond the LHC
sensitivity.}
\label{fig:Uclluh}
\end{figure}

The combined constraints from $B\to \rho\gamma$ and $B\to \mu^+\mu^-$ on
$\ollup$ and $\ollhp$ are shown in Figure~\ref{fig:Uclluh}. The region between
the dashed lines is beyond the LHC reach, but the LHC will be able to exclude
(though perhaps not completely) the non-SM region in Fig.~\ref{fig:Uclluh}. In
the case of $\ollup$ and $\orlwp$ the present data are not strong enough to
exclude the non-SM region allowed by $B\to \rho \gamma$.

\subsection{\boldmath $\Delta F=2$ transitions}
\label{sec:mixing}

In this section we present the results of the analysis of the NP effects in
$\Delta F=2$ processes. Since their contribution appear at the same time in $B_d
\ov B{}_d$, $B_s \ov B{}_s$ and $K^0 \ov K{}^0$, we performed a full fit using
the CKMFitter code \cite{ckmfitter}, after having suitably modified it to
include the results of Sec.~\ref{sec:df2}. The code simultaneously fits
experimental data for the Wolfenstein parameters $\ov\rho$ and $\ov\eta$ and for
NP (extending earlier studies in $\Delta F=2$
processes~\cite{Ligeti:2004ak,dms}). The observables used here include the $B_d$
and $B_s$ mass differences, the time dependent $CP$ asymmetries in $B\to
J/\psi K$, the $CP$ asymmetries in $B\to \pi\pi,\, \rho\rho,\, \rho\pi$, the
ratio of $|V_{ub}|$ and $|V_{cb}|$ measured in semileptonic $B$ decays, the $CP$
asymmetries in $B\to D K$, the width difference in the $B_s \ov B{}_s$ system,
$\Delta\Gamma_s$, the semileptonic $CP$ asymmetry in $B$ decays, $A_{\rm SL}$,
and the indirect $CP$ violation in $K$ decays, $\epsilon_K$. We allowed the NP
Wilson coefficients to be complex and performed a scan over their phases. Thus,
the results in this section are quoted in terms of the absolute values of the
$C_i$ and $C'_i$.  Keeping only one operator at a time, we get
\begin{eqnarray}
|\cllu| &<& 0.07\,, \qquad |\cllh| < 0.014\,, \qquad |\crlw| < 0.14\,, \nn\\*
|\cllup| &<& 0.11\,, \qquad |\cllhp| < 0.018\,, \qquad |\crlwp| < 0.26\,.
\end{eqnarray}

As before, we also performed a combined analysis for the $LL$ operators. This is
particularly interesting for $\ollup$ and $\ollhp$, since until $B\to X_d \ell^+
\ell^-$ data becomes available, only $\Delta F=2$ processes are sensitive to the
complex phases. In general, allowing for a variation of the phases of $\cllup$
and $\cllhp$, a cancellation can occur between the two contributions and the
above bounds are relaxed.  If their absolute values satisfy $|\cllhp| \sim 0.1\,
|\cllup|$ then arbitrarily large values of the Wilson coefficients is allowed
for some values of the phases. This possibility is ruled out when the $B\to \rho
\gamma$ and $B\to \mu^+\mu^-$ constraints are included. Indeed, combining
$\Delta F=2$ with these measurements, we obtain
\beq
|\cllup| < 0.26\,, \qquad |\cllhp| < 0.026\,.
\eeq

\section{Combined constraints and conclusion}
\label{conclusion}

In this paper, we studied constraints on flavor-changing neutral current top
quark decays, $t\to cZ,\, uZ,\, c\gamma,\, u\gamma$.  We used an effective field
theory in which beyond the SM physics is integrated out.  In the theory with
unbroken electroweak symmetry the leading contributions to such FCNC top decays
come from seven dimension-6 operators of Eq.~(\ref{operators}).  We assumed that
the new physics scale, $\Lambda$, is sufficiently above the electroweak scale,
$v$, to expand in $v/\Lambda$ and neglect higher dimension operators.  We find
different and sometimes stronger constraints than starting with an effective
theory which ignores $SU(2)_L$ invariance.

The 95\% CL constraints on the Wilson coefficients of the operators involving
3rd and 2nd generation fields are summarized in Table~\ref{tab:Cconstraints}. 
We consider one operator at a time, i.e., that there are no cancellations. The
top two rows show the present direct constraints and the expected LHC bounds. 
The next three rows show the bounds from $B$ physics. In the $B\to X_s\gamma,\,
X_s\ell^+\ell^-$ row the combined bounds from these processes are shown.  The
two allowed regions are obtained neglecting the complex phases of the operators
(see Fig.~\ref{fig:clluhmix} and the discussion in Sec.~\ref{sec:bsgll}). This
assumption can be relaxed in the future with more detailed data on $B\to
X_s\ell^+\ell^-$.  In the $\Delta F=2$ row the numbers refer to upper bounds on
the magnitudes of the Wilson coefficients and are derived allowing the phase to
vary.  The best bound for each operator is listed and then translated to a lower
bound on the scale $\Lambda$ (in TeV, assuming that the $C$'s are unity), and to
the maximal $t\to c Z$ and $t\to c \gamma$ branching ratios still allowed by
each operator. The last row indicates whether a positive LHC signal could be
explained by each of the operators alone.  In this row, the star in
``Closed$^*$" for $\cllu$ and $\cllh$ refers to the fact that small values of
these Wilson coefficients cannot give an observable top FCNC signal, however,
there is an allowed region with cancellations between the SM and the NP, which
may still give a signal.  In the same row ``Ajar" means that $\crlw$ and $\crlb$
cannot yield an LHC signal in $t\to cZ$ but may manifest themselves in $t\to
c\gamma$. It is remarkable that the coefficients of several operators are better
constrained by $B$ physics than by FCNC top decays at the LHC.

\begin{table}[t]
\begin{tabular}{c|c|c|c|c|c|c|c}
	&  $\cllu$  &  $\cllh$  &  $\crlw$  &  $\crlb$  
	&  $\clrw$  &  $\clrb$  &  $\crru$ \\
\hline\hline
direct bound  &  9.0  &  9.0  &  6.3  &  6.3  &  6.3  &  6.3  &  9.0 \cr
LHC sensitivity	&  0.20  &  0.20  &  0.15  &  0.15  &  0.15  &  0.15  &  0.20 \cr
\hline
$B\to X_s \gamma,\ X_s \ell^+\ell^-$  &  [$-$0.07, 0.036]  
  &  $\matrix{[-0.017,\, -0.01] \cr [-0.005,\, 0.003]}$  &  [$-$0.09,\,0.18]  
  &  [$-$0.12,\,0.24]  &  [$-$14,\,7]  &  [$-$10,\,19]  &  --- \cr
$\Delta F = 2$	&  0.07  &  0.014  &  0.14  &  ---  &  ---  &  ---  &  --- \cr
semileptonic	&  ---  &  ---  &  ---  &  ---  &  [0.3,\,1.7]  &  ---  &  --- \cr
best bound	&  0.07  &  0.014  &  0.15  &  0.24  &  1.7  &  6.3 &  9.0  \cr
\hline
$\Lambda$ for $C_i=1$ (min)  &  3.9\tev   &  8.3\tev  &  2.6\tev  & 2.0\tev  
  &  0.8\tev  &  0.4\tev &  0.3\tev \cr
${\cal B} (t\to c Z)$ (max)
  &  7.1\,$\times 10^{-6}$  &  3.5\,$\times 10^{-7}$  &  3.4\,$\times 10^{-5}$
  &  8.4\,$\times 10^{-6}$  &  4.5\,$\times 10^{-3}$  &  5.6\,$\times 10^{-3}$
  &  0.14 \cr
${\cal B} (t\to c \gamma)$ (max)
  &  ---  &  ---  &  1.8\,$\times 10^{-5}$  &  4.8\,$\times 10^{-5}$  
  &  2.3\,$\times 10^{-3}$  &  3.2\,$\times 10^{-2}$  &  ---  \cr
\hline\hline
LHC Window  & Closed$^*$  & Closed$^*$  & Ajar  &  Ajar  &  Open  &  Open  &  Open\\
\hline\hline
\end{tabular}
\caption{95\% CL constraints on the Wilson coefficients of the operators
involving 3rd and 2nd generation fields for $\Lambda = 1\tev$.  The top two rows
show the present direct constraints and the expected LHC bounds. The second part
shows the bounds from $B$ physics, which is then translated to a lower bound on
the NP scale, $\Lambda$, and to the maximal $t\to c Z$ and $t\to c \gamma$
branching ratios each operator could still give rise to (the ATLAS sensitivity
with 100\,fb$^{-1}$ is $5.5\times 10^{-5}$ and $1.2\times 10^{-5}$,
respectively). The last line concludes whether a positive LHC signal could be
explained by each of the operators.}
\label{tab:Cconstraints}
\end{table}

Table~\ref{tab:Uconstraints} shows the constraints on the operators involving
the 3rd and 1st generation quarks.  We studied this because the LHC may not be
able to distinguish between $t\to c$ and $t\to u$ FCNC decays, and these
processes are also interesting in their own rights.   In this case there are two
allowed regions of $\clrwp$ from semileptonic decays, as can be seen in
Fig.~\ref{fig:btou-incexcl}.  The entries in the ``combined bound" row show the
result of the fit to all the $B$ decay data above it, as discussed in
Sec.~\ref{sec:mixing}. We see from the last row that the LHC window remains open
for all of the $RR$, $LR$, and $RL$ operators, except $\orlwp$.

We conclude from Tables~\ref{tab:Cconstraints} and \ref{tab:Uconstraints} that
if the LHC sees FCNC $t$ decays then they must have come from $LR$ or $RR$
operators, unless there are cancellations.  Moreover, if $t\to cZ$ is seen but
$t\to c\gamma$ is not, then only $\orru$ could account for the data.

\begin{table}[t]
\begin{tabular}{c|c|c|c|c|c|c|c}
	&  $\cllup$   &  $\cllhp$   &  $\crlwp$  &  $\crlbp$  
	&  $\clrwp$  &  $\clrbp$  &  $\crrup$ \\
\hline\hline
direct bound	&  9.0	&  9.0  &  2.7  &  2.7  &  2.7  &  2.7  &  9.0 \cr
LHC sensitivity	&  0.20  &  0.20  &  0.15  &  0.15  &  0.15  &  0.15  &  0.20 \cr
\hline
$B\to \rho \gamma,\ \mu^+\mu^-$  &  $\matrix{[-0.26,\, -0.21] \cr [-0.026,\, 0.03]}$
  &  [$-$0.023,\,0.026]  &  $\matrix{[-0.7,\, -0.5] \cr [-0.07,\, 0.08]}$  
  &  $\matrix{[-0.1,\, 0.09] \cr [0.7,\, 0.9]}$  &  ---  &  ---  &  --- \cr
$\Delta F = 2$	&  0.11  &  0.02  &  0.26  &  ---  &  ---  &  ---  &  --- \cr
semileptonic	&  ---  &  ---  &  ---  &  ---  &  $\matrix{[-0.9,\, 0.1] \cr [0.8,\, 1.4]}$  &  ---  &  --- \cr
combined bound	&  0.10  &  0.02  &  0.16  &  0.9  &  1.4  &  2.7  &  9.0 \cr
\hline
$\Lambda$ for $C_i=1$ (min)  &  3.2\tev  &  7.2\tev  &  2.5\tev  &  1.1\tev
  &  0.8\tev  &  0.6\tev  &  0.3\tev \cr
${\cal B} (t\to u Z)$ (max)
  &  1.6\,$\times 10^{-5}$  &  6.4\,$\times 10^{-7}$  &  4.1\,$\times 10^{-5}$
  &  1.2\,$\times 10^{-4}$  &  3.2\,$\times 10^{-3}$  &  1.0\,$\times 10^{-3}$
  &  0.14 \cr
${\cal B} (t\to u \gamma)$ (max)
  &  ---  &  ---  &  2.1\,$\times 10^{-5}$  &  6.7\,$\times 10^{-4}$  
  &  1.6\,$\times 10^{-3}$  &  5.9\,$\times 10^{-3}$  &  ---  \cr
\hline\hline
LHC Window  &  Closed  &  Closed  &  Ajar  &  Open  &  Open  &  Open  &  Open \\
\hline\hline
\end{tabular}
\caption{Constraints on the Wilson coefficients of the operators involving 3rd
and 1st generation fields.  The entries in the table are analogous to
Table~\ref{tab:Cconstraints}.}
\label{tab:Uconstraints}
\end{table}

Our analysis used the currently available data and compared it to an estimate of
the LHC reach with 100\,fb$^{-1}$.  However, by that time many of the
constraints discussed above will improve, and new measurements will become
available.  The direct bounds will be improved by measurements from Run~II of
the Tevatron in the near future. All the $B$ decay data considered in this paper
will improve, and the calculations for many of them may become more precise. 
Important ones (in no particular order) are: (i)~improved measurements of $B\to
X_s\ell^+\ell^-$ to better constrain the magnitudes and especially the phases of
$\cllu$, $\cllh$, $\crlw$, and $\crlb$; (ii)~measurements of the lepton energy
and the hadronic mass moments in $B\to X_u\ell\bar\nu$ to constrain $\clrwp$;
(iii) improvements in $B\to \rho\gamma$ and measurement of $B\to X_d\gamma$ to
reduce the uncertainties of $\cllup$, $\cllhp$, $\crlwp$, and $\crlbp$; (iv)
measurement of and $B\to X_d\ell^+\ell^-$ to reduce the errors and constrain the
weak phases of these last four coefficients.  Additional information will also
come from LHCb. For example, the measurement of the $CP$ violating parameter
$S_{B_s\to \psi\phi}$, the direct measurements of the CKM angle $\gamma$, and
some of the above rare decays will help improve the constraints.  With several
of these measurements available, one can try to relax the no-cancellation
assumption employed throughout our analysis. Note that not all NP-sensitive $B$
factory measurements can be connected to FCNC top decays; e.g., the $CP$
asymmetry $S_{K^*\gamma}$ is sensitive to right-handed currents in the down
sector and cannot receive a sizable enhancement from the operators in
Eq.~(\ref{operators}).  Thus, there are many ways in which there can be
interesting interplays between measurements of or bounds on FCNC $t$ and $b$
quark decays.

If an FCNC top decay signal is observed at the LHC, the next question will be
how to learn more about the underlying physics responsible for it.  With a few
tens of events one can start to do an angular analysis or study an integrated
polarization asymmetry~\cite{Agashe:2006hk}.  These could discriminate
left-handed or right-handed operators (say $\orru$ or $\ollu$).  Such
interactions could arise in models in which the top sector has a large coupling
to a new physics sector, predominantly through right-handed
couplings~\cite{tcR}. However, a full angular analysis that could also
distinguish $\orru$ from $\olrw$ requires large statistics, which is probably
beyond the reach of the LHC. 

The observation of FCNC top decays at the LHC would be a clear discovery of new
physics, and therefore it would be extremely exciting.  Our analysis shows that
an LHC signal requires $\Lambda$ to be less than a few TeV. This generically
implies the presence of new particles with significant coupling to the top
sector. If the new particles are colored, we expect that they will be discovered
at the LHC.  It would be gratifying to decipher the underlying structure of new
physics from simultaneous information from top and bottom quark decays and
direct observations of new heavy particles at the LHC.

\acknowledgments

We acknowledge helpful discussions with Chris Arnesen and Iain Stewart on
$B\to\pi\ell\bar\nu$~\cite{Arnesen:2005ez}, Frank Tackmann on $B\to
X_s\ell^+\ell^-$~\cite{Lee:2006gs}, and Phillip Urquijo about the Belle
measurement of $B\to X_c\ell\bar\nu$~\cite{Urquijo:2006wd}.
P.F., M.P., and G.P.\ thank the Aspen Center for Physics for hospitality while
part of this work was completed.
The work of P.F., Z.L., and M.P.\ was supported in part by the  Director, Office
of Science, Office of High Energy Physics of the  U.S.\ Department of Energy
under contract DE-AC02-05CH11231.
M.S.\ was supported in part by the National Science Foundation under grant
NSF-PHY-0401513.

\appendix

\section{Analytic expressions}

\def\Wslash{W\!\!\!\!\!\slash\,\,}
\def\Zslash{Z\!\!\!\!\slash\,}

We give the form of the operators of Eq.~(\ref{operators}) after electroweak
symmetry breaking, keeping only trilinear vertices which do not involve the
Higgs:
\begin{eqnarray}
\ollu &=& \frac{\sqrt{2}m_W^2}{g_2} \left(\overline{t}_L \Wslash^- s_L
  + \overline{b}_L\Wslash^+ c_L \right) + \frac{2m_Z m_W}{g_2}\,
  \overline{t}_L\Zslash c_L + \ldots
\,,\nn\\
\ollh &=& \frac{2m_Z m_W}{g_2} \left(\overline{t}_L\Zslash c_L 
  + \overline{b}_L\Zslash s_L\right)+ \ldots \,, \nn\\
\orlw &=& m_W\overline{s}_L \sigma^{\mu\nu}t_R W_{\mu\nu}^- + 
\sqrt{2}m_W \overline{c}_L \sigma^{\mu\nu} t_R \left( c_w Z_{\mu\nu} 
  + s_w A_{\mu\nu} \right) + \ldots
\,, \nn\\
\orlb &=& \sqrt{2} m_W \overline{c}_L \sigma^{\mu\nu} t_R \left( s_w A_{\mu\nu}
  - \frac{s_w^2}{c_w}\, Z_{\mu\nu} \right) + \ldots\,, \nn\\
\olrw &=& m_W\overline{b}_L  \sigma^{\mu\nu}c_R W_{\mu\nu}^-
  + \sqrt{2}m_W \overline{t}_L \sigma^{\mu\nu} c_R \left( c_w Z_{\mu\nu} 
  + s_w A_{\mu\nu} \right) + \ldots\,, \nn\\
\olrb &=& \sqrt{2} m_W  \overline{t}_L \sigma^{\mu\nu} c_R \left( s_w A_{\mu\nu}
  - \frac{s_w^2}{c_w}\, Z_{\mu\nu} \right) + \ldots\,, \nn\\
\orru &=& \frac{2m_Z m_W}{g_2}\, \overline{t}_R\Zslash c_R + \ldots \,.
\end{eqnarray}
Here $s_w = \sin\theta_w$, $c_w = \cos\theta_w$, and the dots denote hermitian
conjugate and the neglected vertices involving Higgs and higher number of
fields.  Throughout this paper the covariant derivative is
defined as $D_\mu = \partial_\mu + i g A^a_\mu \tau^a + i g^\prime B_\mu
$.

The analytic expressions for the contributions of the operators in
Eq.~(\ref{operators}) to the top FCNC partial widths are
\begin{eqnarray}\label{rates}
{\Gamma}(t\to c Z)
&=& \frac{m_{t}}{16 \pi}\frac{v^{2}m_{t}^{2}}{\Lambda^{4}}\, (1-y)^{2} 
  \Big\{\left[| \cllh  + \cllu |^2 + | \crru |^2 \right](1+2y) \nonumber\\
&&{} + 2g_{2}^{2}\cos^{2}\theta_{W}(2+y) \left[|\clrb \tan^2 \theta_W-\clrw| 
^2+ | \crlb \tan^2 \theta_W-\crlw|^2\right] \nn\\
&&{} - 6 \sqrt{2} g_{2} \sin\theta_W \tan\theta_W \sqrt{y}\, \mathrm{Re}
   \left[(\crlb)^*(\cllh + \cllu) - (\clrb)^*\crru\right] \nn\\
&&{} + 6\sqrt{2} g_{2} \cos\theta_W \sqrt{y}\, \mathrm{Re}
   \left[(\crlw)^* (\cllh + \cllu) - (\clrw)^* \crru\right] \Big\}\,, \\
{\Gamma}(t\to c \gamma) &=&
   \alpha m_{t} \frac{v^2 m_{t}^{2}}{\Lambda^4}
   \left( | \clrb + \clrw |^2 + | \crlb + \crlw |^2
   \right)\,,
\end{eqnarray}
where $y=m_{Z}^{2}/m_{t}^{2}$. The analogous expressions for $t\to u$ decays are
obtained by replacing $C_i$ by $C'_i$ above. This expression makes it
straightforward to relate the Wilson coefficients used in this paper with
different notation present in the literature, which defines the couplings in the
effective Lagrangian after electroweak symmetry breaking.

Next we present the analytic expression for the Wilson coefficients originating
from the operators in Eq.~(\ref{operators}).  We use the
$\overline{{\mathrm{MS}}}$ scheme and match at the scale $\mu = m_W$. It is
easiest to express the results as modifying the Inami-Lim functions $B_0$,
$C_0$, and $D_0$/$D_0'$, coming from box diagrams, $Z$-penguins, and
$\gamma$-penguins, respectively.  Using the standard normalization of the
effective Hamiltonian
\begin{equation}\label{hnorm}
{\cal H}_{\mathrm{eff}} = - \frac{G_F}{\sqrt2} V_{tb}V_{ts}^* \sum C_i\, O_i\,,
\end{equation}
the Wilson coefficients at the matching scale can be written as
\begin{eqnarray}
C_{7 \gamma} &=& - \frac{1}{2} D_0' (x) \,, \nn\\*
C_{9 V} &=& \frac{\alpha}{2 \pi} \left[- \frac{1}{\sin^2 \theta_W} B_0(x) + 
  \left(\frac{1}{\sin^2 \theta_W} - 4\right) C_0(x) - D_0(x)\right] , \nn\\*
C_{10 A} &=& \frac{\alpha}{2 \pi} \frac{1}{\sin^2 \theta_W} 
  \big[B_0(x) - C_0(x) \big] , 
\end{eqnarray}
where $x = m_t^2/m_W^2$. In the SM, we have the well-known
expressions~\cite{BBL}
\begin{eqnarray}\label{BCDSM}
B_0 (x) &=& \frac{1}{4} \left[ - \frac{x}{x-1} 
  + \frac{x}{(x-1)^2} \ln x \right] , \nn\\*
C_0 (x) &=& \frac{x}{8} \left[ \frac{x-6}{x-1} 
  + \frac{3x+2}{(x-1)^2} \ln x \right] , \nn\\*
D_0 (x) &=& - \frac{4}{9} \ln x - \frac{19 x^3 - 25 x^2}{36(x-1)^3} 
  + \frac{x^2 (5 x^2 - 2 x - 6)}{18 (x-1)^4} \ln x \,,\nn\\*
D'_0 (x) &=& \frac{8 x^3 + 5 x^2 - 7x}{12(x-1)^3} 
  - \frac{3 x^3 - 2 x^2}{2 (x-1)^4} \ln x \,.
\end{eqnarray}

The contributions of the $\ollu, \olrw$, and $\olrb$ operators introduced in
Eq.~(\ref{operators}) can be included by adding the following terms to
Eq.~(\ref{BCDSM})
\begin{eqnarray}
\Delta B_0 (x) &=& \frac{\kappa}{2}\, \cllu \left(\frac{1}{x-1}
  - \frac{x\ln x}{(x-1)^2}\right) \,,
\\
\Delta C_0 &=& \frac{\kappa}{24}\, \cllu 
  \left(\frac{20(x-1) \sin^2\theta_W+23 x+7}{x-1}
  - \frac{6 x (x^2+x+3)}{(x-1)^2} \ln x \right) \nn\\*
&&{} - \frac{2\pi \kappa}{\alpha_{2}}\, \cllh 
  + \frac{3\kappa g}{2\sqrt{2}}\, \crlw\, \sqrt x \left(\frac{x}{x-1}
  - \frac{x\ln x}{(x-1)^2}\right) 
  - \frac{\kappa \sqrt{x\, x_c}}8\, \crru  \left(\frac12 
  - \frac{x-4}{x-1}\, \ln x \right) , \label{DC0}
\\
\Delta D_0 &=& - \frac{\kappa}{9}\, \cllu \left(\frac{47x^3-237x^2+312x-104}
  {6(x-1)^3} - \frac{3x^4-30x^3+54x^2-32x+8}{(x-1)^4} \ln x \right) \nn\\*
&&{} + \frac{\sqrt2\, \kappa g}{3}\, \crlw\,
  \sqrt{x}\left(\frac{49x^2-89x+34}{6(x-1)^3}
  - \frac{6x^3-9x^2+x+1}{(x-1)^4}\ln x\right)
  - \sqrt2\, \kappa g\, \crlb \frac{\sqrt{x}\, \ln x}{x-1} \nn\\
&&{} + \frac{ \kappa g\sqrt{x_c}}{\sqrt2}\, \clrw \left( \frac{59 x-68}{9(x-1)}
  + \frac{3x-2}{(x-1)^2} \ln x \right) 
  + \frac{\kappa g \sqrt{x_c}}{\sqrt{2}}\, \clrb\, \frac{x-2}{x-1} \,,
\\
\Delta  D_0' &=& \frac{\kappa}{2}\, \cllu \left(
  \frac{68x^3-291x^2+297x-92}{18(x-1)^3}
  + \frac{x^2(3x-2)}{(x-1)^4} \ln x \right)
  + \frac{4\kappa}{27}\, \cllh (\sin^2\theta_W + 3) \nn\\*
&&{} - \frac{\kappa g}{3\sqrt2}\, \crlw \sqrt{x}\left(\frac{3x^3+33x^2-25x+1}
  {2(x-1)^3} - \frac{3x^4-6x^3+33x^2-32x+8}{(x-1)^4} \ln x \right)\nn\\
&&{} + \frac{\kappa g}{2\sqrt{2}}\, \crlb \sqrt{x}
  \left(\frac{x-7}{x-1} - \frac{2x(x-4)}{(x-1)^2} \ln x \right) 
  + \frac{2\sqrt2\, \kappa g\sqrt{x_c}}{3}\, \clrw 
  - \frac{\kappa g \sqrt{x_c}}{\sqrt2}\, \clrb \,,
\end{eqnarray}
where $x_c = m_c^2/m_W^2$ and
\begin{equation}
\kappa =  \frac{v^2}{\Lambda^2}\, \frac{V_{cs}^*}{V_{ts}^*} \,.
\end{equation}
Note that the contribution of $\ollh$ to $\Delta C_0$ occurs at tree level, as
indicated by its $1/\alpha_2$ enhancement in Eq.~(\ref{DC0}),
so $\ollh$ gives tree-level contributions to $C_{9V}$ and $C_{10A}$.
Nevertheless, we shall not include the matrix element of $\ollh$ to
one higher order in $\alpha_2$, in analogy with the conventional approach in
which the NNLL calculation of $B\to X_s\ell^+\ell^-$ does not include the ${\cal
O}(\alpha_s^2)$ matrix element of $O_{9V}$.

Finally we calcuate the $\Delta F=2$ contributions due to
$\cllu$ and $\clrw$. The shift in the SM contributions read
\begin{equation}
S_0^{\rm SM}\to S_0^{\rm SM} + \kappa_{i}\, \Delta S_{i}(x) 
  + \kappa_{i}\, \kappa_{j} \Delta S'_{i,j}(x) 
  + \kappa_{ij}\, \Delta S''_{ij}(x) \,,
\end{equation}
where $i=u,h,w$ labels the contributions from the operators $\ollu$, $\ollh$ and
$\orlw$, respectively.  The expressions for $\Delta S$ and $\Delta S'$ are
\begin{eqnarray}
\Delta S_u &=& -\frac{x(4 x^2-11x+1)}{(x-1)^2}
  + \frac{2 x (x^3-6x+2)}{(x-1)^3} \ln x \,,
\\
\Delta S_{h} &=& -\frac{ x [(1+x)\sin^2 \theta_W+2x-6]}{x-1}
  + \frac{2 x [x (x+2)\sin^2\theta_W - 6]}{3(x-1)^2} \ln x \,,
\\
\Delta S_{w} &=& 3\, g\, \sqrt{2x} \left[\frac{
  x (x+1)}{(x-1)^2}-\frac{2 x^2}{(x-1)^3} \ln x\right] ,
\\
\Delta S_{u,u}' &=& \frac{7 x^3-15x^2+6x-4}{(x-1)^2}
  - \frac{2 x (2 x^3+3x^2-12x+4) }{(x-1)^3}\ln x \,,
\\
\Delta S_{h,h}' &=& \frac{16\pi}{\alpha_2} \,,
\\
\Delta S_{u,h}' &=& \frac{2 [(x+1)(x+2) \sin^2\theta_W 
  + 3(x^2-9x+4)]}{3(x-1)} 
  + \frac{2x [x (x-3-2 \sin^2 \theta_W)+6]}{(x-1)^2} \ln x \,,
\\
\Delta S_{w,w}' &=& g^2 \left[ -\frac{6 x(x+1)}{(x-1)^2}
  + \frac{12 x^2}{(x-1)^3} \ln x \right] ,
\end{eqnarray}
and $\kappa_i$ depends on the flavor transition,
\begin{equation}\label{kappas}
\kappa_i = \frac{v^2}{\Lambda^2}  \left\{
\begin{array}{ll}
C_i\, V_{cs}/V_{ts}  & {\rm\ for } \ \ t\to c \textrm{ contribution in } b \to
s \,,\\
C_i\, V_{cd}/V_{td}  & {\rm\ for } \ \ t\to c \textrm{ contribution in } b \to
d \,,\\
C'_i\, V_{us}/V_{ts}  & {\rm\ for } \ \ t\to u \textrm{ contribution in } b \to
s \,,\\
C'_i\, V_{ud}/V_{td}  & {\rm\ for } \ \ t\to u \textrm{ contribution in } b \to
d \,,\\
(C_i\, V_{ts}^*V_{cd} + C_i^*\, V_{cs}^*V_{td}) / (V_{ts}^*V_{td})  & {\rm\ for } \
\ t\to c \textrm{ contribution in } s \to d \,,\\
(C'_i\, V_{ts}^*V_{ud} + C_i^{\prime*}\, V_{us}^*V_{td}) / (V_{ts}^*V_{td}) & {\rm\ for } \
\ t\to u \textrm{ contribution in } s \to d \,.\\
\end{array}
\right.
\end{equation}
The $\kappa_{ij}$ are zero except for $K^0\ov K{}^0$ mixing, where they are
given by
\begin{equation}
\kappa_{ij} =\frac{v^4}{\Lambda^4} \left\{
\begin{array}{cc}
C_i C_j^*\, V_{cs}^*V_{cd}/(V_{ts}^*V_{td})
& {\rm\ for } \ \ t\to c,\\
C'_i C_j^{\prime*}\, V_{us}^*V_{ud}/(V_{ts}^*V_{td}) & {\rm\ for } \ \ t\to u,\\
\end{array}
\right.
\end{equation}
and $\Delta S''_{ij}$ are given by
\begin{eqnarray}
\Delta S_{u,u}'' &=&   \frac{x (29 x^2-84x+7)}{4(x-1)^2}
  - \frac{ x (7 x^3+9x^2-64x+24)}{2(x-1)^3}\ln x \,,
\\
\Delta S_{u,h}'' &=&\frac{2x [2x-6+(x+1)\sin^2\theta_W]}{(x-1)}
  - \frac{4x [x(x+2) \sin^2\theta_W - 6]}{3(x-1)^2} \ln x \,,
\\
\Delta S_{w,w}'' &=& g^2 \left[ -\frac{2x(x^2-2x-11)}{(x-1)^2}
  - \frac{12x^2 (x^2-3x+4)}{(x-1)^3} \ln x \right] .
\end{eqnarray}

\end{document}